\documentclass[aps,prd,12point,nofootinbib,twocolumn,superscriptaddress]{revtex4-2}

\usepackage{subfigure}
\usepackage{color}
\usepackage{mathrsfs}
\usepackage{graphicx}
\usepackage{amssymb, bm}
\usepackage{amsmath, amsthm}
\usepackage{epstopdf}
\usepackage{hyperref}
\usepackage{enumerate}
\usepackage{longtable}

\newcommand{\be}{\begin{equation}}
\newcommand{\ee}{\end{equation}}

\begin{document}

\title{Thermodynamics of scalar--tensor gravity}

\author{Valerio Faraoni}
\email[]{vfaraoni@ubishops.ca}
\affiliation{Department of Physics \& Astronomy, Bishop's University, 
2600 College Street, Sherbrooke, Qu\'ebec, 
Canada J1M~1Z7 }

\author{Andrea Giusti}
\email[]{agiusti@phys.ethz.ch}
\affiliation{Institute for Theoretical Physics, ETH Zurich,
Wolfgang-Pauli-Strasse 27, 8093, Zurich, Switzerland}



\begin{abstract} Previously, the Einstein equation has been described as 
an equation of state, general relativity as the equilibrium state of 
gravity, and $f({\cal R})$ gravity as a non-equilibrium one. We apply 
Eckart's first order thermodynamics to the effective dissipative fluid 
describing scalar-tensor gravity. Surprisingly, we obtain simple 
expressions for the effective heat flux, ``temperature of gravity'', shear 
and bulk viscosity, and entropy density, plus a generalized Fourier law in 
a consistent Eckart thermodynamical picture. Well-defined notions of 
temperature and approach to equilibrium, missing in the current 
thermodynamics of spacetime scenarios, naturally emerge.

\end{abstract}

\maketitle

\section{Introduction}
\label{sec:1}

The idea that there is a deep connection between 
gravity and thermodynamics, originating in black hole thermodynamics, took 
a new meaning with Jacobson's seminal work \cite{Jacobson:1995ab} in which 
the Einstein field equation of general relativity (GR) was derived as an 
equation of state using purely thermodynamic considerations. This fact has 
deep implications for viewing gravity as an emergent, rather than 
fundamental, phenomenon and for quantum gravity as well. In this 
picture, quantizing the Einstein equation would make no more sense than 
quantizing the macroscopic ideal gas equation of state, which cannot 
produce 
fundamental results such as the energy spectrum and 
eigenfunctions of the hydrogen atom. In quantum gravity, the ``atoms of 
spacetime'' (if they exist) would have to be found with a radically 
different approach.

A second, equally important, idea was proposed in 
Ref.~\cite{Eling:2006aw}, which derived the field equation of fourth order 
metric $f({\cal R}) $ gravity using thermodynamics. This modification of 
GR would correspond to dissipative, non-equilibrium ``thermodynamics of 
gravitational theories'' in which a ``bulk viscosity of spacetime'' is 
introduced to explain dissipation, while GR corresponds to equilibrium 
thermodynamics instead. These works have generated a very large 
literature. In particular, Ref.~\cite{Chirco:2010sw} stressed the 
essential role of shear viscosity while eliminating bulk viscosity from 
this picture. In spite of the large literature, the equations ruling how 
modified gravity approaches the GR equilibrium state remain a mystery, and 
the order parameter (for example, the temperature) regulating this 
dissipative phenomenon has not been identified.

Here we propose an approach to the last two problems in the spirit of the 
above-mentioned ideas, but in a different context. We consider the large 
class of scalar-tensor theories of gravity \cite{BransDicke,ST} (which 
contains the $f({\cal R}) $ subclass), a minimal modification of GR in 
which a scalar degree of freedom $\phi$ is added to the usual two massless 
spin two modes of GR. The contribution of $\phi$ to the field equations is 
described as an effective relativistic dissipative fluid 
\cite{Pimentel89,Faraoni:2018qdr}. We then apply Eckart's first order 
thermodynamics \cite{Eckart40} to this effective fluid and extract 
explicit expressions for the relevant effective thermodynamic quantities, 
including the heat current density, ``temperature of modified gravity'', 
viscosity coefficients, and entropy density.

To summarize our results, the temperature is positive-definite and 
vanishes at the GR equilibrium state; the bulk viscosity is absent, and 
the shear viscosity coefficient is negative, which allows for the 
possibility that the entropy density decreases, in agreement with the fact 
that the system (the $\phi$-fluid) is not isolated. What is more, we 
provide an equation describing explicitly the approach of scalar-tensor 
gravity to the GR equilibrium state, which is Eckart's generalization of 
the Fourier law modelling diffusion \cite{Eckart40}.

Begin with the scalar-tensor action \footnote{We follow the 
notation of Ref.~\cite{Waldbook} and we use units in which Newton's 
constant $G$ and the speed of light $c$ are unity.} 
\be
S_\text{ST} =  \int d^4x \frac{\sqrt{-g}}{16\pi}  \left[ \phi {\cal R} 
-\frac{\omega(\phi )}{\phi} 
\, \nabla^c\phi \nabla_c\phi -V(\phi) \right] +S^\text{(m)} \,,  
\label{STaction}
\ee
where ${\cal R}$ is the Ricci scalar, the Brans-Dicke scalar $\phi>0$ is 
approximately the inverse of the 
effective gravitational 
coupling, $\omega(\phi)$ is the ``Brans-Dicke coupling'', $V(\phi)$ is a 
potential, and $S^\text{(m)}=\int d^4x \sqrt{-g} \, 
{\cal  L}^\text{(m)} $ is the  
matter action. The corresponding  field equations \cite{BransDicke, ST} 
are written as the effective Einstein equations  
\begin{eqnarray}
{\cal R}_{ab} - \frac{1}{2}\, g_{ab} {\cal R} &=& \frac{8\pi}{\phi} \,  
T_{ab}^\text{(m)}   + \frac{\omega}{\phi^2} \left( \nabla_a \phi 
\nabla_b \phi -\frac{1}{2} \, g_{ab} 
\nabla_c \phi \nabla^c \phi \right) \nonumber\\
&&\nonumber\\
&\, &  +\frac{1}{\phi} \left( \nabla_a \nabla_b \phi 
- g_{ab} \Box \phi \right) 
-\frac{V}{2\phi}\, 
g_{ab} \,, \label{BDfe1} 
\end{eqnarray}
\be
\Box \phi = \frac{1}{2\omega+3}   \left( 
\frac{8\pi T^\text{(m)} }{\phi}   + \phi \, \frac{d V}{d\phi} 
-2V -\frac{d\omega}{d\phi} \nabla^c \phi \nabla_c \phi \right) \,, 
\label{BDfe2}
\ee
where ${\cal R}_{ab}$ is the Ricci tensor and $ T^\text{(m)} $  is the 
trace of the matter stress-energy tensor $T_{ab}^\text{(m)} $. The 
terms containing $\phi$ and its derivatives form an effective 
$\phi$-fluid with 
stress-energy tensor 
\begin{eqnarray}
&&8\pi T_{ab}^{(\phi)} = \frac{\omega}{\phi^2} \left( \nabla_a \phi 
\nabla_b \phi - 
 \frac{1}{2} \, g_{ab} \nabla^c \phi \nabla_c \phi  \right) \nonumber\\
&& + 
 \frac{1}{\phi} \left( \nabla_a \nabla_b \phi -g_{ab} \square \phi \right) 
- \frac{V}{2 \phi} \, g_{ab} \,. \label{BDemt}
\end{eqnarray}

\section{Effective scalar field fluid}
\label{sec:2}


When the gradient $\nabla^a \phi$ is timelike, it is used to 
construct the normalized effective  fluid 4-velocity  
\be
u^a  = \frac{\nabla^a  \phi}{\sqrt{ -\nabla^e \phi \nabla_e \phi }} \,.
\label{4-velocity}
\ee
The $3+1$ splitting of spacetime into the time direction $u^c$ and the 
3-dimensional space of the observers comoving with the fluid follows. 
Their 3-space is endowed with the 
Riemannian metric $h_{ab} \equiv g_{ab} + u_a u_b $
and  ${h_a}^b$ is the projection operator on this 3-space. The 
effective fluid 4-acceleration  $ \dot{u}^a \equiv u^b \nabla_b 
u^a $ is orthogonal to the 4-velocity. The spatial projection of the 
velocity gradient  
\be
V_{ab} \equiv  {h_a}^c \, {h_b}^d \, \nabla_d u_c 
= \theta_{ab}= \sigma_{ab} +\frac{\theta}{3} \, h_{ab}\,,
\ee 
coincides with the symmetric expansion tensor $\theta_{ab}$ since 
its antisymmetric part (the vorticity $\omega_{ab}$) vanishes, as  
$u^a$ derives from a  gradient. $u^a$ is irrotational and  
hypersurface-orthogonal \cite{Ellis71, Waldbook}.  Here $\theta \equiv 
{\theta^c}_c =\nabla_c 
u^c$, while 
$ \sigma_{ab} \equiv \theta_{ab}-\theta\, h_{ab}/3 $ is the 
symmetric, trace-free shear tensor. The velocity gradient splits as 
\cite{Ellis71}  
\be
\nabla_b u_a =  \sigma_{ab}+\frac{\theta}{3} \, h_{ab} -  \dot{u}_a 
u_b  \,. 
\label{ecce}
\ee
When these general definitions \cite{Ellis71, Waldbook} are 
specialized to the effective $\phi$-fluid, one obtains 
\cite{Faraoni:2018qdr}
\begin{widetext}
\begin{eqnarray}
\dot{u}_a &=& \left( -\nabla^e \phi \nabla_e \phi \right)^{-2} 
\nabla^b \phi 
\Big[ (-\nabla^e \phi  \nabla_e \phi)  \nabla_a \nabla_b 
\phi + \nabla^c  \phi \nabla_b \nabla_c \phi \nabla_a \phi \Big] \,, 
\label{acceleration}\\
&&\nonumber\\
\theta &=& 
\frac{ \square  \phi}{ \left (-\nabla^e \phi 
\nabla_e \phi \right)^{1/2} }  + \frac{ \nabla_a 
\nabla_b \phi \nabla^a \phi \nabla^b \phi }{ \left( -\nabla^e \phi 
\nabla_e 
\phi \right)^{3/2} } \,, \label{thetaScalar}\\
&&\nonumber\\
\sigma_{ab} &=&  \left( -\nabla^e \phi \nabla_e \phi \right)^{-3/2} \left[ 
-\left( \nabla^e  \phi \nabla_e 
\phi \right) \nabla_a \nabla_b  \phi 
- \frac{1}{3} \left(  \nabla_a \phi \nabla_b \phi   - g_{ab} \, \nabla^c 
\phi \nabla_c \phi   \right) \square \phi  \right.\nonumber\\
     &&\nonumber\\ 
     &\, & \left. - \frac{1}{3} \left( g_{ab} + \frac{ 2 \nabla_a \phi 
\nabla_b \phi }{   \nabla^e \phi \nabla_e \phi } 
 \right) \nabla_c \nabla_d 
\phi \nabla^d \phi \nabla^c \phi + \left( \nabla_a \phi \nabla_c 
\nabla_b 
\phi + \nabla_b \phi \nabla_c \nabla_a \phi \right) \nabla^c \phi \right]  
\,. \label{sheartensor}
\end{eqnarray} 
The effective stress-energy tensor~(\ref{BDemt}) of the 
Brans-Dicke-like field takes the imperfect fluid form 
\be 
T_{ab} = \rho u_a u_b + q_a u_b + q_b u_a + \Pi_{ab} 
\,, \quad\quad \quad\quad \Pi_{ab}= P  h_{ab} +\pi_{ab} 
\label{imperfectTab}
\ee 
with effective energy density, heat flux density, stress tensor, 
isotropic pressure, and anisotropic stresses 
\cite{Pimentel89,Faraoni:2018qdr} 
\begin{eqnarray}
8 \pi \rho^{(\phi)} &=&  -\frac{\omega}{2\phi^2} \, \nabla^e \phi \nabla_e 
\phi  +  \frac{V}{2\phi} + \frac{1}{\phi} \left( \square \phi -  
\frac{  \nabla^a \phi \nabla^b \phi \nabla_a 
\nabla_b \phi}{ \nabla^e \phi  \nabla_e \phi  } \right)  
\,,\label{effdensity}\\
&&\nonumber\\
8 \pi q_a^{(\phi)}   &=& \frac{\nabla^c  \phi \nabla^d \phi}{\phi 
  \left(-\nabla^e \phi \nabla_e \phi \right)^{3/2} } \,  
\Big(  \nabla_d \phi \nabla_c \nabla_a \phi 
- \nabla_a \phi \nabla_c \nabla_d \phi \Big) \,, \label{eq:q}\\
&&\nonumber\\
8 \pi \Pi_{ab}^{(\phi)}  &=&  
 \left( -\frac{\omega}{2\phi^2} \, \nabla^c \phi \nabla_c \phi 
-\frac{\Box\phi}{\phi} -\frac{V}{2\phi} \right) h_{ab} +\frac{1}{\phi} \, 
{h_a}^c {h_b}^d \nabla_c \nabla_d \phi \,,  \label{eq:effPi2}\\
&&\nonumber\\
8 \pi P^{(\phi)}  & = &  - \frac{\omega}{2\phi^2} \, \nabla^e \phi 
\nabla_e \phi - 
\frac{V}{2\phi} - \frac{1}{3\phi}  \left( 2\square \phi + 
\frac{\nabla^a \phi \nabla^b \phi \nabla_b \nabla_a \phi }{\nabla^e \phi 
\nabla_e  \phi }  \right) \,, \label{effpressure}\\
&&\nonumber\\
8 \pi \pi_{ab}^{(\phi)}   &=& \frac{1}{\phi \nabla^e \phi \nabla_e 
\phi } 
\left[ \frac{1}{3} \left( \nabla_a  \phi \nabla_b \phi - g_{ab} \nabla^c 
\phi \nabla_c \phi \right) \left(  \square \phi  - 
\frac{ \nabla^c \phi  \nabla^d \phi \nabla_d \nabla_c \phi }{ \nabla^e \phi 
\nabla_e \phi }   
\right) \right. \nonumber\\
&&\nonumber\\
&\, & \left. + \nabla^d \phi \left(  \nabla_d \phi \nabla_a \nabla_b 
\phi - 
\nabla_b \phi \nabla_a \nabla_d  \phi - \nabla_a \phi \nabla_d \nabla_b 
\phi +  
\frac{ \nabla_a \phi \nabla_b \phi  \nabla^c \phi \nabla_c 
\nabla_d \phi }{ \nabla^e \phi \nabla_e \phi } \right) \right] \,. 
\label{piab-phi}
\end{eqnarray} 
\end{widetext}

\section{Eckart's thermodynamics for scalar-tensor gravity}
\label{sec:3}

Since it is a fact that the field equations of scalar-tensor 
gravity assume the form of effective Einstein equations with an effective 
imperfect fluid (plus ordinary matter) as source, it makes sense to take 
this property further and examine the thermodynamical context of this 
imperfect fluid, which is Eckart's first order thermodynamics. The 
non-causal spacelike heat flow finds its place in Eckart's theory which, 
albeit non-causal, is widely used as a first approach to relativistic 
thermodynamics.

In Eckart's first order thermodynamics (\cite{Eckart40}, see also 
\cite{Andersson:2006nr}), the 
dissipative quantities 
(viscous pressure $P_\text{vis}$, heat current density $q^c$, and 
anisotropic 
stresses $\pi_{ab}$) are related to the expansion $\theta$, temperature 
${\cal T}$, and shear 
tensor $\sigma_{ab}$ by the constitutive equations \cite{Eckart40}
\begin{eqnarray}
P_\text{vis} &=& -\zeta \, \theta \,,\\
&&\nonumber\\
q_a &=& -K \left( h_{ab} \nabla^b {\cal T} + {\cal T} \dot{u}_a \right) 
\,, \label{Eckart}\\
&&\nonumber\\
\pi_{ab} &=& - 2\eta \, \sigma_{ab} \,,\label{def:eta}
\end{eqnarray}
where $\zeta$ is the bulk viscosity, $K$ is the thermal conductivity, and 
$\eta$ is the shear viscosity. The comparison of Eqs.~(\ref{eq:q}) and 
(\ref{acceleration}) yields \cite{Faraoni:2018qdr} 
\be
q_a^{(\phi)} =  -\frac{ \sqrt{-\nabla^c \phi \nabla_c \phi}}{ 8 \pi \phi} \, 
\dot{u}_a \,.\label{q-a}  
\ee 
In the  comoving frame, the  spatial temperature gradient vanishes 
identically and the heat flow arises solely from the inertia of energy.  
The Eckart temperature of the $\phi$-fluid, which can be called the 
``temperature of 
scalar-tensor gravity'', is then \cite{Faraoni:2018qdr}
\be
{\cal T}= \frac{ \sqrt{-\nabla^c \phi \nabla_c\phi}}{8\pi K \phi} \,;
\label{temperature}
\ee
it is positive definite and vanishes when $\phi=$~const., which 
corresponds to GR.

The structure of $T_{ab}^{(\phi)}$  does not allow 
for bulk viscosity, hence $\zeta=0$. Comparing Eqs.~(\ref{piab-phi}) 
and~(\ref{sheartensor}) for  $\pi_{ab}^{(\phi)}$ and 
$\sigma_{ab}^{(\phi)}$ and using
Eq.~(\ref{def:eta}) yields
\be
\eta= - \frac{ \sqrt{-\nabla^c \phi \nabla_c \phi}}{16\pi \phi} 
= -\frac{K{\cal T}}{2} < 0 \,.
\ee
Negative viscosities are common in fluid mechanics (including, {\em e.g.},  
atmospheric  physics, ocean currents, liquid crystals) in the 
presence of turbulence  and appear in non-isolated   
systems into which energy is fed from the outside ({\em e.g.}, 
\cite{negviscosity}). Indeed, the $\phi$-fluid is not isolated. In the 
action~(\ref{STaction}), $\phi $ couples explicitly to gravity through the 
term $\phi {\cal R}$.

In the different context of spacetime thermodynamics, 
Ref.~\cite{Chirco:2010sw} stressed the importance of shear viscosity and 
the absence of bulk viscosity in $f({\cal R})$ gravity, contrary to the 
previous interpretation of \cite{Eling:2006aw}. These results are echoed 
in our approach.

In Eckart's formalism, the entropy current due to the heat flux is 
$R^a=q^a/{\cal T}$ \cite{Eckart40,Andersson:2006nr}, 
with components $\left( 0, \vec{q} /  {\cal T}  \right) $  in the comoving 
frame. The entropy current density in a fluid with particle density $n$ 
and entropy density $s$ is $
s^a = sn u^a +R^a $, where $R^a$ (here equal to $ -K\dot{u}^a $) describes 
entropy generation due to  dissipative processes. 
While, in a  non-dissipative fluid, entropy is conserved ($\nabla_c 
s^c=0$), with dissipation in an isolated system it is  $\nabla_c s^c >0$ 
due to $R^a$.
 
Using Eqs.~(\ref{effdensity}), (\ref{effpressure}), and 
(\ref{temperature}), the entropy density obtained from the first law of 
thermodynamics is 
\begin{eqnarray}
s &\equiv & \frac{dS}{dV}= \frac{\rho+P}{{\cal T}} =
 \frac{K}{ \sqrt{-\nabla^e \phi \nabla_e \phi}} \nonumber\\
&&\nonumber\\
&\, & \times \left[ 
-\frac{\omega}{\phi} \, \nabla^e\phi \nabla_e \phi +\frac{\Box\phi}{3} 
-\frac{4}{3} \, \frac{ \nabla^a\phi \nabla^b \phi \nabla_a\nabla_b \phi}{ 
\nabla^e\phi \nabla_e \phi} \right]\, , \nonumber\\
&&
\end{eqnarray}
assuming a closed (yet, not isolated) system. Furthermore,
in a fluid in which the particle number is conserved, $\nabla_a n^a=0$ 
(where $n^a = nu^a$ is the particle current density), one has 
\cite{Eckart40, Andersson:2006nr}
\be
\label{eq:sa}
\nabla_c s^c =  \frac{ P_\text{vis}^2}{\zeta {\cal T}} +\frac{ q_c q^c}{ 
K{\cal 
T}^2} + \frac{\pi_{ab} \pi^{ab}}{2\eta {\cal T}} \,,
\ee
where the bulk viscosity term is 
absent for the effective $\phi$-fluid. 
The comparison of the components of the effective stress-energy tensor of 
the $\phi$-fluid based on Eckart's constitutive laws imply 
Eqs.~\eqref{def:eta}, \eqref{q-a}, \eqref{temperature} and a 
vanishing 
contribution of the bulk viscosity term, then Eq. \eqref{eq:sa} reduces to

\be
\nabla_c s^c = K\left( \dot{u}^a \dot{u}_a + \frac{ K {\cal 
T} \sigma^2}{\eta} \right) = K\left( \dot{u}^a \dot{u}_a 
-\sigma_{ab}\sigma^{ab} \right) \,,\label{entropyincrease}
\ee
where $\sigma^2 \equiv \sigma_{ab}\sigma^{ab}/2$.     
Since the second term in round brackets is negative, the entropy 
does not always increase.\footnote{If they 
are possible, situations in which the $\phi$-fluid is geodesic, 
$\dot{u}^a=0$, correspond to decreasing entropy density, consistent with 
the fact that then $R^a=0$  and shear contributes 
to decreasing $s$ due to the negative $\eta$, as 
described by Eq.~(\ref{entropyincrease}).} Indeed, if energy is injected 
into the $\phi$-fluid coupled to gravity, $s$ can 
decrease.

\section{The approach to the GR equilibrium state}
\label{sec:4}

An effective heat equation for the $\phi$-fluid, governing the approach to 
equilibrium, can easily be obtained by differentiating the quantity $K 
\mathcal{T}$ in Eq. \eqref{temperature},  which yields
\be
\frac{d\left( K{\cal T}\right)}{d\tau} = 8\pi \left( K{\cal T}\right)^2 
-\theta\, K{\cal T} +\frac{\Box\phi}{ \sqrt{-\nabla^e\phi \nabla_e\phi}} 
\, , \label{HeatEq}
\ee
which differs from the standard 
result of Eckart's first-order thermodynamics 
\cite{Eckart40,Andersson:2006nr}.

The physical interpretation of Eq.~(\ref{HeatEq}) is rather tricky, 
however one can gain some insight into the approach to equilibrium 
of the system by considering simplified scenarios. First, let 
us consider electrovacuum, $\omega=$~const., and $V(\phi) = 
0$;  the field equations then imply that $\Box\phi = 0$. Therefore, if $\theta 
< 0$, then
\be
\frac{d( K{\cal T} )}{d\tau} > 8\pi (K{\cal T})^2 \, ,
\ee
meaning that $K{\cal T}$ diverges away from the 
GR equilibrium state extremely fast. Thus, one can easily observe that 
near spacetime singularities, where the wordlines of the $\phi$ field 
converge, the deviations of scalar-tensor gravity from GR will be extreme. 
Such a scenario is therefore worthy of further investigation in relation 
with analytic solutions of scalar-tensor theories 
involving naked singularities (see {\em e.g.}, \cite{reviewbello}). As a 
second scenario, let us consider again the same electrovacuum 
with  $\theta > 0$. Given these assumptions one has that, in principle, 
the second term in the right-hand side of Eq.~(\ref{HeatEq}) can dominate 
over the 
first one. This implies that the solution $K{\cal T}$ approaches zero or, 
in other words, the theory approaches the GR equilibrium. One can 
understand this in a rather evocative way observing that the expansion 
seems to cool down gravity. Nonetheless, if $K{\cal T}$ is large, the 
positive term will dominate the right-hand side and drive the solution 
away from GR. Therefore, the approach to the GR equilibrium state is not 
necessarily granted.  Examples of analytic solutions supporting these 
simple arguments will be presented elsewhere.

\section{Conclusions and outlooks}

Contrary to Jacobson's thermodynamics 
of spacetime \cite{Jacobson:1995ab, 
Eling:2006aw}, our approach has minimal assumptions. Rewriting the 
scalar-tensor field equations in the form of Einstein equations with an 
imperfect fluid does not entail any extra assumption. The limitations of 
our approach are those intrinsic to Eckart's first order thermodynamics: 
it is not causal and suffers from instabilities, but is nevertheless the 
most widely used model of relativistic thermodynamics due to its relative 
simplicity. Causality 
violation is also present in the imperfect fluid 
model~(\ref{imperfectTab}) which is also the most common model of 
dissipative fluid. This fact leads us to associate, rather naturally, 
Eckart's 
thermodynamics with the effective fluid description of 
scalar-tensor 
gravity. While, {\em a priori}, Eckart's thermodynamics applied to an 
effective fluid could have led nowhere, it surprisingly makes sense for 
the $\phi$-fluid and one can read expressions of the 
heat current density, effective temperature (the sought-for order 
parameter), shear  viscosity, (vanishing) bulk 
viscosity, entropy density, and the approach to the GR equilibrium state 
is described by Eckart's generalization~(\ref{Eckart})  of the Fourier  
law. 
 
Two unexpected results are that {\em i})~the heat flux is purely inertial 
in the frame comoving with the $\phi$-fluid, in which the spatial gradient 
of ${\cal T}$ vanishes; {\em ii})~ the shear viscosity is negative, 
leading to the possibility of decreasing entropy. This fact is not 
disconcerting because the system is not isolated and negative (turbulent) 
viscosities are common in the fluid dynamics of non-isolated systems. An 
important consequence is that gravity does not necessarily relax to the GR 
state but can depart from it. This fact opens the 
possibility that other states of equilibrium, different from GR, exist.

We stress that the effective fluid formalism is not an analogy, but a 
definite approach to the problem of gravity reaching the GR 
state of equilibrium through dissipation.  Separate publications will 
widen the approach of this Letter\footnote{An alternative approach used, 
thus far, in braided kinetic gravity \cite{Pujolas:2011he} would trade 
temperature with chemical potential and assign zero temperature and 
entropy to the effective fluid.} and will attempt to generalize it to 
causal (second order) thermodynamics to overcome the limitations intrinsic 
to Eckart's formalism.

\begin{acknowledgments} 
 We thank Jeremy C\^ot\'e 
for a discussion. This work is supported, in part, by the Natural Sciences 
\& Engineering Research Council of Canada (Grant~2016-03803 to V.F.). A. 
Giusti is supported by the European Union's Horizon 2020 research and 
innovation programme under the Marie Sk\l{}odowska-Curie Actions (grant 
agreement No. 895648 -- CosmoDEC) and his work was carried out in the 
framework of the activities of the Italian National Group for Mathematical 
Physics [Gruppo Nazionale per la Fisica Matematica (GNFM), Istituto 
Nazionale di Alta Matematica (INdAM)]. \end{acknowledgments}


\end{document}